\documentclass[12pt,preprint]{aastex}
\newcommand\etal{{\rm et~al\/}.}

\newcommand\alwaysmath[1]{\ifmmode{#1}\else{$#1$}\fi}

\newcommand\angstrom{{\,{\rm \AA}}} 
\newcommand\angs{\angstrom}

\newcommand\kel{\alwaysmath{\,{\rm  K}}}

\newcommand\astron{Astronomy}
\newcommand\dept{Dept.~}
\newcommand\phys{Physics}
\newcommand\das{\dept of \astron}

\newcommand\dph{\dept of \phys}
\newcommand\dphas{\dph\ and \astron}
\newcommand\uof{University of~}
\newcommand\aadv{Astrophysical Advances}

\newcommand\uc{\uof California}
\newcommand\uofa{\uof Arizona}

\newcommand\uva{\uof Virginia}

\newcommand\lick{UCO/Lick Observatory}

\newcommand\nasa{National Aeronautics and Space Administration (NASA)}

\newcommand\stew{Steward Observatory}
\newcommand\stsci{Space Telescope Science Institute}
\newcommand\ilc{i}
\newcommand\iilc{ii}
\newcommand\logg{\alwaysmath{ \log\,g }}

\newcommand\teff{\alwaysmath{T_{{\rm eff}}}}

\newcommand\feh{{\rm [Fe/H]}}

\newcommand\alfe{\alwaysmath{\rm [\alpha/Fe]}}
\newcommand\cfe{\alwaysmath{\rm [C/Fe]}}
\newcommand\nfe{\alwaysmath{\rm [N/Fe]}}
\newcommand\ofe{\alwaysmath{\rm [O/Fe]}}
\newcommand\mgfe{\alwaysmath{\rm [Mg/Fe]}}
\newcommand\sife{\alwaysmath{\rm [Si/Fe]}}
\newcommand\cafe{\alwaysmath{\rm [Ca/Fe]}}
\newcommand\tife{\alwaysmath{\rm [Ti/Fe]}}

\def\2h#1{\alwaysmath{{\rm [#1/H]}}}
\def\2fe#1{\alwaysmath{{\rm [#1/Fe]}}}

\newcommand\mgi{{\rm Mg\,{\textsc \ilc}}} 
\newcommand\mgj{{\rm Mg\,{\textsc \iilc}}}

\newcommand\caj{{\rm Ca\,{\textsc \iilc}}}

\newcommand\ebmv{\alwaysmath{E(B-V)}}

\newcommand\vmk{\alwaysmath{V-K}}

\newcommand\Hbeta{\alwaysmath{{\rm H}\beta}} 
\newcommand\Hgamma{\alwaysmath{{\rm H}\gamma}} 
\newcommand\Hdelta{\alwaysmath{{\rm H}\delta}} 
\newcommand\Hepsilon{\alwaysmath{{\rm H}\epsilon}}

\newcommand\ngc{\alwaysmath{\rm NGC\,}}
\newcommand\hd{\alwaysmath{\rm HD\,}}

\newcommand\ocen{\alwaysmath{\omega\,{\rm Cen}}}
\slugcomment{To appear in {\it The Astrophysical Journal}}
\shorttitle{ }
\shortauthors{Peterson et al.}

\begin{document}

\title{Blue Horizontal Branch Stars in Old, Metal-Rich Stellar  
Systems\altaffilmark{1,2}}

\author{
Ruth C. Peterson\altaffilmark{3},
Bruce W. Carney\altaffilmark{4},
Ben Dorman\altaffilmark{5},
Elizabeth M. Green\altaffilmark{6},
Wayne Landsman\altaffilmark{7},
James Liebert\altaffilmark{8},
Robert W. O'Connell\altaffilmark{9}, and
Robert T. Rood\altaffilmark{10}
}

\altaffiltext{1}{Based on observations obtained with the Hubble Space 
Telescope of \stsci, under contract with the \nasa}
\altaffiltext{2}{Based on observations obtained with the Shane Telescope at 
Mt.\ Hamilton, \lick}
\altaffiltext{3}{Ruth C. Peterson,
   \lick, \das, \uc, Santa Cruz, CA 95064, and 
\aadv, Palo Alto, CA  94301; peterson@ucolick.org}
\altaffiltext{4}{Bruce W. Carney,
   \dphas, \uof North Carolina, Chapel Hill, NC  27599-3255;                       bruce@physics.unc.edu}
\altaffiltext{5}{Ben Dorman,
   L-3Com Analytics Corp., Greenbelt, MD 20771;
   Ben.Dorman@gsfc.nasa.gov}
\altaffiltext{6}{Elizabeth M. Green,
   \stew, \uofa, Tucson, AZ 85721-0065; bgreen@as.arizona.edu}
\altaffiltext{7}{Wayne Landsman,
   Science Systems \& Applications, Inc., Greenbelt, MD 20771;
   landsman@mpb.gsfc.nasa.gov}
\altaffiltext{8}{James Liebert,
   \stew, \uofa, Tucson, AZ 85721-0065; liebert@as.arizona.edu}
\altaffiltext{9}{Robert W. O'Connell,
   \das, \uva, P.O. Box 3818, Charlottesville, VA  22903-0818;
   rwo@virginia.edu}
\altaffiltext{10}{Robert T. Rood,
   \das, \uva, P.O. Box 3818, Charlottesville, VA  22903-0818;
   rtr@veris.astro.virginia.edu}

\begin{abstract}

Twenty years ago, Burstein et al.\ (1984) recognized that the
metal-rich globular clusters in the Andromeda galaxy (M31)
exhibited anomalously strong Balmer and CN lines
compared to Milky Way clusters. They suggested younger ages 
might be the cause, unless blue stars above the main-sequence 
turnoff or on the horizontal branch were uncommonly prominent.
Here we test these suggestions 
by fitting the detailed mid-ultraviolet
(2280-3120\angs) and optical (3850-4750\angs) spectra of one moderately
metal-rich M31 globular cluster, G1.  We explore the effects of 
a wide range of non-solar temperatures and abundance ratios, 
by combining a small set of theoretical stellar spectra like those 
of Peterson et al.\ (2001) that were calculated using 
extensively updated atomic-line constants.
To match the mid-UV fluxes of G1, we find that hot components 
with $T_e \geq 8000$\kel\ must be included.
We obtain a very good fit with cool and hot blue horizontal branch (BHB) stars,
but less satisfactory fits for 
blue straggler stars, those hotter than the main-sequence turnoff.

The G1 color-magnitude diagram does show cool BHB 
stars, and the color of its giant branch supports the metallicity
of one-sixth the solar value deduced from the composite spectrum with
BHB stars.  The turnoff temperature of the best-fit model is
consistent with that of turnoff stars in galactic globular clusters and the
field halo, indicating G1 is comparably old. Because metal-rich
cool BHB and extremely blue HB stars have now been found within our own Galaxy --- 
in open clusters, globular clusters, and the field of the Bulge --- we suggest 
that these hot horizontal-branch stars be considered in fitting 
spectra of metal-rich populations such as the Andromeda globular clusters,
to avoid possible underestimates of their ages.
We plan to make the relevant spectral calculations available as part
of our Hubble Treasury Program.
\end{abstract}


%\pagebreak

%\citep[e.g.,][]{roo73}

\section{Introduction}

The old stellar populations in globular clusters and galaxies
are important tracers of the histories of galaxies and 
can help decide among various current 
models of the formation of galaxies. For example, if an elliptical galaxy 
forms in a simple, isolated, dynamical collapse, its stars form primarily 
at early times \citep[e.g.,][]{egg62,ari87}. In contrast, the hierarchical 
assemblage of an elliptical galaxy from smaller pieces, in mergers
accompanied by star formation, leads naturally to a 
substantial intermediate-age population \citep[e.g.,][]{kau93}. 

Although globular clusters and elliptical galaxies harbor similar populations
\citep{baa44}, they do have certain disparate properties that make it 
difficult to establish their relative ages. 
The major complication is the variation in elemental abundances \citep{cha96}.
Luminous elliptical galaxies are metal-rich and show an enhancement in the 
magnesium-to-iron abundance ratio relative to that of the Sun 
\citep[e.g.,][]{oco76,pet76,tra00a,tra00b},
while Andromeda globulars reach higher metallicities than do those of
the Galaxy \citep{vdb69}.
\citet{wor94} has shown that there is an ``age-metallicity degeneracy'', i.e.\
very little difference between the optical spectra of two old systems in
which one has twice the age and three-fourths the metallicity of the other.
The Balmer lines of hydrogen can break this 
degeneracy, but their strengths are also influenced by stars hotter than 
the main-sequence turnoff (MSTO).

\citet{bur84} showed that M31 globular clusters of moderate to high 
metallicity exhibit anomalously strong H$\beta$ and CN line absorption 
compared to those of Milky Way globulars, and that their CN (but not H$\beta$) 
is stronger than in nuclei of elliptical galaxies. They 
considered whether the Balmer-line enhancement might be due to 
anomalously hot stars on the blue horizontal branch (BHB), or to 
unevolved blue stragglers situated on the main sequence blueward of the turnoff.
In the end, they favored the possibility that these M31 globulars might 
be younger than those of the Milky Way, since this might
also account for the anomalous CN absorption.

Subsequent work has confirmed several important distinctions among 
these groups, but has generally not supported substantial differences in age. 
\citet{tri89} confirmed from optical spectra 
that CN is enhanced in the metal-rich M31 globulars 
relative to those of the Milky Way, and again argued for younger ages;
but \citet{bro90} noted that the \Hbeta\ anomalies in M31 globulars 
all but disappeared when \Hbeta\ was plotted against a magnesium index
rather than an iron index.
Mid-UV measurements from the Astro/Ultraviolet Imaging Telescope \citep{boh93} 
suggested that the ages of the overwhelming majority of M31 globulars
were comparable to those of the Milky Way.
While \citet{fan90} deduced from stellar mid-UV spectra 
obtained with the International Ultraviolet Explorer (IUE)
that UV feature distinctions are primarily due to abundance differences,
\citet{pon98} emphasized that feature strengths 
could also be affected by old, evolved subdwarf B/O stars---the
visually faint but very hot ``extreme'' horizontal branch (EHB) stars
responsible for the upturn in far-UV flux seen in luminous ellipticals
\citep{oco99}. 

In this paper we re-examine these issues using the \citet{pon98} spectra 
of the moderately metal-rich M31 globular cluster G1 ($=$ K1 $=$ Mayall II).
This bright, spatially extended, well-studied cluster 
is partially resolved into stars by the 
Hubble Space Telescope (HST) Wide Field/Planetary Camera 2 (WFPC2).
\citet{ric96} found that its $V,~V-I$ color magnitude
diagram (CMD) bore a strong resemblance to that  of the
Milky Way globular cluster 47 Tucanae, suggesting a similar age and
metallicity for both clusters. They also noted a population of 
cool blue horizontal branch (BHB) stars near their detection limit. 
%The subdwarf B (sdB) stars are visually much fainter, and would not have been detected.

With $\feh = -0.7$ \citep{ric96} or $-0.9$ 
\citep{mey01}, G1 has the lowest metallicity of any Andromeda globular 
cluster that shows a sizable ($\sim$40\%) \Hbeta\ enhancement.
%It is sufficiently metal-poor that 
Its spectrum can now 
be analyzed using theoretical mid-UV stellar spectral calculations.
\citet[][PDR01]{pet01b}
%Peterson, Dorman, \& Rood (2001; PDR01) 
have shown that their {\it ab initio}
calculations of spectra from 2280 to 3120\angs, 
incorporating extensively updated constants and pseudo-identifications 
for atomic absorption-line features,
reproduce observed spectra for low-metallicity solar-type stars
up to a metallicity of $\feh = -0.5$ (one-third solar).

In this work, we fit simple composite theoretical spectra to the 
\citet{pon98} G1 mid-UV spectra
(2280-3120\angs, 5.5\angs\ resolution)
and optical spectra (3850-4750\angs, 8.8\angs\ resolution)
taken with the Hubble Space
Telescope (HST) Faint Object Spectrograph (FOS). We use the observed 
CMD's for G1 \citep{ric96} and 47 Tuc \citep{bri97} to guide our
modeling. We forego a detailed spectral synthesis in favor of a weighted 
combination of the spectra of a few representative types of star,
to avoid introducing a large number of free parameters whose individual
effects cannot be judged. Consequently we ignore for now
the possibility of a spread in metallicity among G1 stars \citep{mey01}.

\citet{ros99} and \citet{sch02a,sch02b} have recently 
used empirical spectral libraries to explore fits 
to high-resolution optical and mid-UV spectra 
of old populations. Such fits are necessarily limited by the 
characteristics of the population of relatively nearby, bright stars, 
which includes
very few old hot stars and stars with non-solar abundance ratios. 
In contrast, our theoretical approach allows us to explore explicitly 
the effects of stars with a wide range of abundance ratios, and of 
hot stars representing both young main-sequence A stars
and the hot horizontal-branch stars of the halo population.
Our results illustrate the basic gains achieved 
in fitting integrated optical and UV spectra 
when stellar parameters can be varied at will.
As part of our future work under our
Hubble Treasury program GO-9455, we will eventually produce a library of
theoretical stellar spectra, enabling a detailed spectral
synthesis. However, our primary results from this simple modeling are
robust enough and interesting enough that we want to draw
attention to them at this stage.

\section{Spectral Calculations}

The spectral calculations for this work were run with the \citet{kur81} program SYNTHE following 
the procedures of PDR01. The same models were employed, those of 
\citet{cas97} downloaded from the Kurucz web site\footnotemark\footnotetext{URL: http://cfaku5.harvard.edu}
and modified at the surface, for PDR01 showed that only models such as 
these --- in which convective overshoot in the photosphere is turned off  ---
can simultaneously reproduce the optical and mid-UV spectra of solar-type stars.
The input list of mid-UV line parameters was largely that of PDR01.
We made additional changes to Kurucz transition probabilities for atomic lines 
identified in the laboratory, added even more ``missing'' lines not yet 
so identified, and halved the absorption due to the \mgi\ bound-free edge 
near 2510\angs, to better match Space Telescope 
Imaging Spectrograph (STIS) spectra of a dozen stars of near-solar 
temperature spanning a wide range of metallicity. 
The input list of optical line parameters is based on 
the Kurucz laboratory atomic line parameters, again  
with corrections like those of \citet{pet93}
to match the Sun but without adding missing lines. 

Theoretical molecular line parameters were used 
in both wavelength regions, with modifications to match the Sun. In the mid-UV,
we reduced by 0.15 dex both the theoretical  
OH transition probabilities of Goldman for the 0-0, 1-1, 2-2, 1-0, 2-1, and 3-2
bands, and the Lifbase CH lines, all provided by 
M. Bessell (private communication, Aug.\ and Oct.\ 2000 respectively). 
To similarly normalize the remaining mid-UV bands of 
both molecules, we reduced the Kurucz theoretical transition probabilities 
by $\Delta  = -0.4 - 0.5 \times$ELO, 
where ELO is the lower energy level (in wavenumbers). 
In the optical, the Kurucz CH transition probabilities %(HYDRIDES0450.100) 
were lowered by 
0.15 dex for lines with at least one energy level measured in the 
laboratory, and by 1.0 dex for lines with both energy levels predicted 
rather than measured. The Kurucz CN transition probabilities 
%(CNAX0400.100) 
were used without change.

Composite spectra were generated by coadding spectra calculated for
up to six representative types of star. Their characteristics 
are given in Table~1. Our choice of atmospheric model
parameters for cool stars was based on those listed in Table~1 of \citet{bri97} 
and plotted as a function of \vmk\ in Fig.\ 4 of that paper.
The temperatures and gravities of the MSTO components
were varied within the range shown to give the best fit for a given
coaddition. The temperatures and gravities of the other components
were held fixed.

For a given coaddition, only one metallicity \feh\ and one relative-abundance
distribution was used, except that \feh\ was fixed as described below
for the model of the EHB star.
For the remaining models, we chose \feh\ to fit most metal lines, 
and \alfe\ (= \mgfe, \sife, \cafe, and \tife)
to match the \mgj\ lines at 2800\angs\ and 
\caj\ H + \Hepsilon\ and \caj\ K.
The relative abundance of CNO elements 
was dependent on the model. Because of the strong CN 
index, giant models incorporated $\ofe = -0.1$, $\cfe = -0.3$, and
$\nfe = +0.4$, while other models adopted $\ofe = 0.1$ and $\cfe =
\nfe = 0.0$. 

To form the composite spectrum, we chose relative flux-weighting factors
for each model as described below. At each wavelength we multiplied the
flux-weighting factor for each model by its flux per unit surface area
as calculated by SYNTHE, and summed the results. The resulting composite
spectrum was Gaussian smoothed and multiplied by a normalization
constant, one for the mid-UV and one for the optical, to match
the observations. These constants
should be equal if %reddening is absent, or if
the observations have been properly dereddened and normalized.

\clearpage
%\begin{table}

\begin{deluxetable}{lccccc} 
\tablewidth{460pt}
\label{tab:starparms} 
\tablecaption{ Representative Stellar Types and their Relative Numbers in Models} 
\tablehead{ 
 & & & \colhead{Mass} & \multicolumn{2}{c}{\underline {Relative Numbers in Models:}}\\
\colhead{Type of Star} & 
\colhead{$T_{\rm eff}/K$} & 
\colhead{$\log\,g$} & 
\colhead{($M_\odot$)} &
\colhead{BHB Models} &
\colhead{BSS Models}\\
}
\startdata 
Main Sequence & 5750 & 4.5 & 0.80 & 1463--1502 & 1186\\
Turnoff       & 5750--6500 & 4.2--4.3 & 0.80 & 923--948 & 748\\
Red Giant     & 4250       & 1.5 & 0.65 & 22 & 22\\
Red HB        & 4500       & 2.0 & 0.60 & 77 & 77\\
F5: Cool Blue Straggler & 6750 & 4.2 & 1.00 &    & 317\\
A7: Warm Blue Straggler & 8000 & 3.8 & 1.40 &    & 225\\
A0: Hot Blue Straggler & 10000 & 3.7 & 2.00 &    & 2--5\\
Cool BHB       & 8000      & 3.0 & 0.55 & 7--11\\
Extreme HB    & 26000      & 5.5 & 0.55 & 12\\
\enddata 
\end{deluxetable} 
%\end{table}

\clearpage

\begin{deluxetable}{ccccccccccccccc}
\rotate
\tabletypesize{\small}
\tablewidth{620pt}
\label{tab:modparms}
\tablecaption{Characteristics of Models Used for the Plots of Theoretical Composite Spectra}
\tablehead{
\multicolumn{2}{c}{Plot Position} &
\colhead{Turnoff} & & & 
\multicolumn{9}{c}{Flux Weights} & Flux\\
\cline{1-2} \cline{6-14}
\colhead{Fig.\ 1} &
\colhead{Fig.\ 2} &
\colhead{\teff (K)} & 
\colhead{\feh} & 
\colhead{\alfe} & 
\colhead{EHB} & 
\colhead{BHB} & 
\colhead{A0} & 
\colhead{A7} &
\colhead{F5} & 
\colhead{TO} & 
\colhead{MS} & 
\colhead{RHB} & 
\colhead{RGB} & 
\colhead{Ratio}\\
}
\startdata
Top & & 6000 & $-0.8$ & 0.2 & 0.000020 & 0.006 & ... & ... & ... & 0.037 & 0.037 & 0.46 & 0.46 & 1.019\\
& Top & 6100 & $-0.8$ & 0.2 & 0.000022 & 0.004 & ... & ... & ... & 0.038 & 0.038 & 0.46 & 0.46 & 0.980\\
2nd & & 5900 & $-0.9$ & 0.2 & 0.000020 & 0.006 & ... & ... & ... & 0.037 & 0.037 & 0.46 & 0.46 & 1.041\\
& 2nd & 6250 & $-0.5$ & 0.2 & 0.000022 & 0.004 & ... & ... & ... & 0.038 & 0.038 & 0.46 & 0.46 & 1.062\\
3rd & & 6500 & $-0.5$ & 0.0 & ... & ... & 0.0008 & ... & 0.02 & 0.030 & 0.030 & 0.46 & 0.46 & 0.948\\
& 3rd & 6500 & $-0.5$ & 0.0 & ... & ... & ...    & 0.05 & ... & 0.037 & 0.038 & 0.46 & 0.46 & 1.030\\
4th & & 6000 & $-0.8$ & 0.2 & ... & 0.006 & ... & ... & ... & 0.037 & 0.037 & 0.46 & 0.46 & 1.101\\
& 4th & 6500 & $-0.5$ & 0.0 & ... & ... & 0.0020 & ... & ... & 0.039 & 0.039 & 0.46 & 0.46 & 0.974\\
5th & & 6250 & $-0.8$ & 0.2 & ... & ... & ... & ... & ... & 0.080 & 0.080 & 0.42 & 0.42 & 0.826\\
& 5th & 6500 & $-0.5$ & 0.0 & ... & ... & ... & ... & ... & 0.080 & 0.080 & 0.42 & 0.42 & 0.830\\
\enddata
\end{deluxetable}

\clearpage

The flux-weighting factors are proportional to
the number of stars represented by a model times the radius squared of the 
stellar atmosphere. The radius may be judged from the stellar mass and 
the model surface gravity \logg. 
Included in Table~1 are the relative numbers of stars thus calculated from 
the flux-weighting factors for each type of model.

Equal flux-weighting factors were adopted for the giant and the RHB
models, and for the MSTO and the main-sequence (MS) stars,
based on approximate numbers of stars in the respective 
regions of the 47 Tuc CMD.  The ratio of the flux-weighting factors 
for the two groups --- giants versus main-sequence stars --- was then 
chosen to best match the slope of the optical spectrum of G1. 

Because the coadded spectra involving just these four types of star
underrepresented the mid-UV flux level, we then added hotter components,
either BHB stars or blue straggler stars (BSS).
Observations of Galactic globular clusters suggest that BHB and/or BSS stars
could exist in G1, yet there is little in the way of
observational or theoretical guidance to suggest how many.
We first tried cool BHB stars, with \teff\ $\sim$8000\kel, and BSS stars
with temperatures moderately hotter than the turnoff. Since the mid-UV 
flux level was still underrepresented,
we added their hotter counterparts, adopting flux-weighting factors that
minimized the discrepancy below 2600\angs\ 
while preserving the slope of the optical continuum.

Temperatures of the hot stars were guided by cluster observations.
The hottest blue stragglers in clusters usually have masses below twice that of 
the cluster turnoff mass \citep[e.g.,][Fig.\ 1]
{fer02}\footnotemark\footnotetext{URL: http://www.astro.virginia.edu/$\tilde{~}$rtr/papers/}.
Our adopted hot BSS model is an extreme case,
an A0V star with a mass twice that of the F5V
star represented by the coolest BSS model \citep{cox99}.
Horizontal branches of globular clusters often are exclusively blue;
some reach temperatures above 20,000\kel, those of EHB stars. 
Our hottest BHB model represents a cool EHB. Using the Kurucz 
ATLAS12 program allowed us to adopt iron-peak abundances one-third solar 
and helium 1/500 solar, to account for radiative levitation and diffusion 
in atmospheres of stars hotter than $\sim$11,000\kel\ \citep[e.g.,][]{gla89,jas90}.

\section{Comparison with Observations} 

In Figures~1 and 2 we show comparisons of ten theoretical composite
spectra (light lines) with observed spectra of G1 (heavy line)
and, at the bottom of each, the comparison of a single stellar model with 
observations of a single star, HD~106516.
The left panel of each figure shows
the mid-UV, and the right, the optical. Each panel shows six
plots, offset vertically; tick marks are separated
by 10\% of the maximum theoretical value plotted in each comparison.
The plots in both figures are
organized vertically according to the presence of hot stars in the 
composite theoretical spectra.  Those of 
the top two plots have both cool BHB and EHB stars.
The third plot of Fig.\ 1 includes two BSS models instead of two BHB models, 
and the third plot of Fig.\ 2 and the fourth plot of both figures includes 
but a single hot-star model.
The fifth plot shows models incorporating MSTO and cooler stars only. 

Table~2 lists the specifics of each composite model spectrum plotted, including 
its position in the figures, the turnoff model temperature, \feh, 
the light-element enhancement \alfe,
the flux-weighting factors, and the ratio of flux normalization constants 
for the composite spectrum. Provided the reddening correction is reliable, 
the flux ratio should be unity to $\sim$5\%, the estimated joint uncertainty 
of the continuum normalizations.

The G1 observations are those of \citet{pon98}, 
as processed by the FOS pipeline on 11 Aug 1994, 
corrected for a reddening \ebmv\ = 0.06\,mag \citep{ric96,mey01} 
via the {\tt deredden} task of 
IRAF\footnotemark\footnotetext{IRAF is written and supported by the 
IRAF programming group at the National Optical Astronomy Observatories (NOAO) 
in Tucson, Arizona. NOAO is operated by the Association of Universities 
for Research in Astronomy (AURA), Inc., under cooperative agreement with the 
National Science Foundation. URL: http://iraf.noao.edu/}, 
and smoothed with a 2\angs\ Gaussian kernel. 
%with the curve of \citet{car89}.
For \hd 106516, the optical 
spectrum is a smoothed version of the continuum-normalized echelle
spectrum PDR01 used to determine these parameters.
It is truncated at 4000\angs, below which line blending renders
continuum definition highly unreliable. Its mid-UV spectrum,
also used by PDR01, is a smoothed echelle spectrum obtained
in Hubble program GO-7433. 
No correction was made for reddening, which should 
be negligible given the star's high latitude
(51$^{\circ}$) and its proximity (23~pc). 

The bottom plot compares observed and calculated spectra for \hd 106516,
using the model parameters from PDR01: \teff\ = 6250\kel, 
\logg\ = 4.3, and \feh\ = $-0.65$, 
with \mgfe\ = +0.25 and \sife\ = \cafe\ = \tife\ = +0.2. 
This model spectrum 
agrees reasonably well throughout 
the mid-UV and in the region redward of 4000\angs. The Balmer lines 
\Hdelta\ (at 4101\angs) and \Hgamma\ (at 4340\,\AA) are well matched, to within
the uncertainties in the continuum definition. However, 
the CH band head near 4306\angs\ is too strong in the calculations.
This probably reflects residual problems with the CH transition probabilities; 
if so, CH will be overestimated in all comparisons. 
Continuum definition affects \hd 106516 alone, for 
the G1 spectra were obtained from space and are on an absolute flux scale. 

Despite its simplicity, our approach leads to very good fits to both
the mid-UV and the optical observations of G1.
Our best-fit spectra are shown at the top of Figures 1 and 2.
Both incorporate cool BHB
stars and EHB stars but not BSS. 
Their iron abundance
\feh\ = $-0.8$, the mild enhancement of light $\alpha$-elements, 
and average turnoff temperature
\teff\ = 6050\,K, all resemble those of mildly metal-poor field halo
and globular cluster stars in the Galaxy.
\hd 106516 is hotter than 6000\,K, but \citet{car01}
identify it as a blue straggler from its binary orbit.  Their Fig.\ 1
shows that it is slightly bluer than field turnoff stars of similar
\feh.  Thus this fit implies a similar age and light-element enhancement
for G1 as for the oldest Galactic stars, except for its CN.

Throughout the mid-UV in the top plot of Fig.\ 1, 
the model fits the G1 spectrum 
nearly as well as the stellar model fits the \hd 106516 spectrum, 
and has very similar discrepancies.
These are attributed
by PDR01 to problems remaining with the mid-UV line list.
In the optical, most features are rather well matched, including
\Hdelta, \caj\ H + \Hepsilon, and \caj\ K.  
(Unfortunately,
the M31 G1 optical spectrum does not extend redward as far as
\Hbeta, at 4861\angs.) 
The CH band head is too strong by roughly the same relative amount 
as in \hd 106516, as is expected if gf-values are in error.
However, the CN band near 3886\angs\ is far too weak.
In \hd 106516 neither this CN band nor the one at 4210\angs\
is detected in the PDR01 echelle spectrum.
Like others before us, we cannot explain the very strong CN in G1.

At the top of Fig.\ 2, we show composite spectra 
designed to test for an age-BHB degeneracy. They are constructed with the same
metallicity and flux-weighting factors as for the best fit, except
that the turnoff temperature was increased and the weight of
the cool BHB model was decreased. This fit is hardly distinguishable 
from the best fit; 
the \caj\ H + \Hepsilon\ and the cores of lines below 2600\angs\ become 
slightly stronger. 
This suggests
that when cool BHB stars are present in signficant but unknown numbers, 
the turnoff temperature 
may be uncertain by $\sim$100\,K even if metallicity and 
reddening are fixed. 
Only the flux ratio rules out a larger uncertainty in temperature and thus age.

The next-to-top plots illustrate the extent to which the age-metallicity 
degeneracy persists in the mid-UV. Both depict fits to BHB+EHB models:
\feh\ = $-0.9$ in Fig.\ 1, and \feh\ = $-0.5$ in Fig.\ 2. Both fits are
still reasonable, marginally worse than the best fit, and both have 
flux ratios of unity to within the uncertainties. The \feh\ = $-0.9$ model 
is unlikely, however, as its turnoff temperature of 5900\,K 
implies an age for the cluster that is probably greater than that 
of the field halo stars. 
A moderately old, mildly metal-poor population remains a possibility for G1. 
This metallicity might be appropriate if abundance rises towards the very center
of G1, since the FOS spectra were taken through a 1\arcsec\ aperture at the 
center. However, even at this metallicity, CN is still not nearly strong enough.

The third plot from the top of Fig.\ 1 portrays composite
spectra with hot and cool blue stragglers instead of EHB and cool BHB stars.
Here, the mid-UV fit is poorer than in the
top plot; neither the \mgj\ region nor the region below 2600\angs\ is as well
reproduced. A similar but more extreme mismatch is seen in 
the third plot from the top of Fig.\ 2, based on 
a single BSS of intermediate temperature. 
A somewhat better fit is shown in the fourth plot from the top of
Fig.\ 2, in which only a hot blue straggler is included. However, 
in all these blue-straggler composite models,
the 2900 -- 3100\angs\ lines are too weak as well. While a higher
metallicity would alleviate this, the optical lines would then be too
strong. Also, the mid-UV magnesium 
lines are poorly matched: the \mgi\ line at 
2856\angs\ is too weak, while the \mgj\ line wings are too weak and its core 
often is too strong. 
Adding main-sequence stars would alleviate this discrepancy,
but again this would spoil the agreement in the optical.
To fit both the mid-UV and optical spectra of G1 thus requires BHB stars 
rather than BSS.

The fourth plot from the top of Fig.\ 1 shows 
a composite spectrum like that of the best fit, but lacking the EHB model.
The optical region is matched rather well, but the observed mid-UV fluxes 
are seriously underestimated in deeper 
lines and at shorter wavelengths. This discrepancy would be reduced 
if scattered light is present in the G1 spectrum. 
However, the IRAF %\footnotemark\footnotetext{URL: http://iraf.noao.edu/}
FOS tasks {\tt countspec} and {\tt bspec} were 
run for a G2V model specifically for this configuration, 
resulting in levels of scattered light 
always less than 3\% for both the mid-UV and the optical G1 spectra.
Only if scattered light comprises $\sim$10\% of the maximum level of 
the mid-UV spectrum would the observed G1 mid-UV spectrum match this 
calculated BHB spectrum without EHB stars.

The fifth plot from the top of both figures portrays composite 
spectra with no hot stars at all. 
In order to preserve the Balmer-line 
strengths and the slope of the optical
continuum at the same metallicity and light-element ratio as the top
plot, the ratio was decreased of RHB + RGB stars to MSTO + MS stars
and the turnoff temperature was raised to 6250K for the plot in Fig.\ 1. 
In the optical, the fit is respectable, although for the Balmer and 
the \caj\ lines the fit is substantially worse than in the top plot.
In the mid-UV, however, the deterioration of the fit is dramatic.
Both line strengths and the slope of the mid-UV flux are very poorly
reproduced. The mid-UV fit is improved but still unsatisfactory in the fifth 
plot of Fig.\ 2, in which a higher metallicity and hotter turnoff temperature 
were adopted.

\section{Discussion}

We have shown that composite theoretical spectra calculated from
scratch with stellar models and our modified line list provide an
excellent match to observed mid-UV and optical spectra of the M31
globular cluster G1, but only if blue horizontal branch stars are
included.  The average turnoff temperature of 6050\kel\ of the best-fit
composite spectra (shown at the top of Figs.\ 1 and 2) is like that of Milky
Way globular clusters and the halo field, suggesting a similarly old age for
G1.  The observed great strength of CN remains unexplained, however.

The metallicity of the best-fit spectrum, $\feh = -0.8$ (one-sixth
solar), agrees well with the \citet{ric96} conclusion from the color
of its giants that G1 is ``at least as metal-rich as 47 Tuc.''
High-resolution spectral analyses of 47 Tuc giants yield $\feh = -0.7$
or $-0.8$ \citep{gra86,bro92,nor95,car97}.  \citet{mey01} infer $\feh
= -0.95$ for G1, but our best fit at $-0.9$ 
requires a turnoff temperature that would render G1 older than Milky
Way globular clusters or the halo field. 
A multi-metallicity model might overcome this objection; 
we have not yet considered such models.

The weighting factors for the best-fit spectra imply as many or more
EHB as cool BHB stars, and one cool BHB star for
every seven to eleven RHB stars in G1---whose CMD otherwise resembles that of 47
Tuc.  The cool-BHB/RHB ratio is consistent with the small but nonzero
number of cool BHB stars seen near the limiting magnitude of the
G1 CMD \citep{ric96}. The EHB stars are visually too faint to have 
been detected there. However, high EHB/cool-BHB ratios are seen in a few
Galactic clusters, notably \ngc 6791 \citep{lie94} and 
$\omega$ Centauri. Although the latter is metal-poor,
its giants clearly span a range of metallicity \citep{nor96},
and its BHB stars appear to have higher metallicities at bluer colors 
\citep{dcr00}.

In the best-fit model of Fig.\ 1, the BHB stars
are significant contributors of UV light. The cool BHB 
stars contribute over one-third of the light below 2550\angs,
$\sim$25\% at 2800\angs, and 20\% near 3100\angs.
The MSTO stars' contribution is 20 -- 25\% below 2550\angs,
and rises to one-third above 2650\angs.
The MS stars contribute about 15\% below 2650\angs, 
and one-quarter of the light over 2800 -- 3100\angs.
Giants plus RHB stars contribute less than
5\% below 2900\angs, rising to 12\% over 2900 -- 3100\angs.
The EHB stars account for one-third of the light below 2400\angs\ but 
25\% at 2550\angs, and $<$10\% above 2900\angs.
The effect of the 
EHB model is to weaken metal line strengths below 2500\angs\
while simultaneously raising the overall fluxes towards the blue end of the
mid-UV. 

The contributions to the 47 Tuc spectrum that \citet{ros99} deduced 
at 2680\angs\ are 15\% from giants and RHB stars and
78\% from MSTO stars, plus 7\% from blue stragglers.
At this wavelength, MSTO and cooler MS stars 
in our best-fit model of Fig.\ 1 contribute 35\% and 24\%, 
versus 3.5\% from giants 
and RHB's, 24\% from cool BHB stars, and 13\% from EHB stars.
In the best-fit model of Fig.\ 2 this becomes 40\% and 25\% for 
MSTO and MS stars, 3.5\% from giants and RHB's, 
16\% from cool BHB's, and 15\% from EHB's.
The large discrepancy in the turnoff/giant ratio between our models 
and that of \citet{ros99} might be due not only to 
our larger hot-star component, but also to the adoption of 
different temperatures and/or abundances of the giant and MS 
stars.
Their choice of empirical templates was based on spectral type, 
which can be misleading at subsolar abundances.

In the optical, 
the EHB stars contribute $\leq$1\% of the light in both our best-fit models.
For that of Fig.\ 1, the MS contributes 14 -- 20\%; the MSTO, 
19 -- 28\%; the cool BHB, $>$25\% below 4000\angs, $\sim$23\% to 
4400\angs, and $\sim$20\% redward; the RHB + RGB, 25\% below 4000\angs,
35\% to 4200\angs, 40\% to 4400\angs, 45\% to 4550\angs, and $>$50\% redward.
Near $\Hbeta$, the RGB + RHB contribution is 55\%; the MS, 14\%; 
the MSTO, 18\%; and the cool BHB, 13\%.

These percentage fluxes and Figs.\ 1 and 2 show that the G1 optical spectrum by itself does 
not demand a hot component. As seen in the next-to-bottom plot in Fig.~1, 
an acceptable fit in the optical 
can be obtained without hot stars, at the same metallicity as the best fit, 
by increasing the proportion of cool stars and raising the turnoff 
temperature by 250\,K -- thereby reducing the inferred age. In Fig.\ 2, 
a similarly good fit is obtained with a higher metallicity and turnoff 
temperature. However, both these fits
are unacceptable in the mid-UV. To match the elevated mid-UV fluxes below 
2650\angs, to fit the shallow cores of strong mid-UV lines, and to preserve 
the strengths of weaker lines at 2900 -- 3100\,\AA\ at the metallicity 
dictated by the optical, models hotter than about 8000\,K are required. 

The effect of such models is illustrated explicitly in Figure~3, 
which plots the theoretical 
fluxes for several models used in generating the composite spectra of
Fig.\ 1. In the optical, the models that are 8000\,K and hotter contribute 
strong Balmer lines but no metallic or molecular lines (except for \caj\ in 
the 8000\,K model), and a continuum that rises towards the blue. 
Together these characteristics explain why the turnoff 
temperature changes when they are included. In the mid-UV, the 8000\,K and 10,000\,K
models show flat, weak-lined spectra, while the 26,000\,K model spectrum 
shows virtually no lines and rises steeply bluewards. In contrast, the models
at lower temperatures show progressively stronger lines and weaker fluxes 
towards the blue.

The fits using model spectra 
with blue stragglers of the same metallicity as the cluster 
are not as good as those with hot horizontal branch stars. This is largely
because the hottest feasible blue straggler model is 10,000\,K, resulting 
in too little flux below 2650\angs\ and too much line dilution near 3000\angs. 
Our discussion of the plots in Figs.\ 1 and 2 has established 
that a BHB population is required to match the G1
spectra, even though our spectral analysis is limited in the 
number of models considered and adopts a constant metallicity.

In previous modeling of old stellar systems, BHB stars have been considered but
usually dismissed. Tools for incorporating them were limited; as described below,
we are planning to address this both theoretically and observationally.
Moreover, there was no theoretical or empirical expectation 
that BHB stars should occur at high metallicity.  
However, cool BHB stars are now known to exist not only in G1, but also 
in several metal-rich environments within the Galaxy.
\citet{ric97} have found BHB stars in two globular clusters in the Galactic
disk, with metallicities one-fourth solar.  In a photometric and
spectroscopic survey for BHB stars in the Galactic bulge,
\citet{pet01a} have found two cool BHB stars with solar metallicity in
a field 7.5$^{\circ}$ from the Galactic center.  The old open
cluster \ngc 6791 contains several possible BHB stars \citep{kal92},
including four or five sdB/O stars (Liebert \etal) %\citep{lie94}.
The coolest of the BHB stars --- both a radial-velocity and a proper-motion
member --- was found by \citet{pet98} to have $\feh = +0.4 \pm 0.1$ dex,
and to have temperature, gravity, and rotational velocity consistent
with a cool BHB star. 

\section{Summary and the Future}

Our theoretical modeling of the spectrum of G1, however simple, has 
achieved very good fits to its observed mid-UV and optical spectra, 
except for CN. 
%The model parameters are all reasonable. 
Only two normalization 
constants were applied to the data; their ratio is consistent 
with the foreground reddening expected for this cluster. The proportions  
of the primary stellar components considered are in accord with observations
of Galactic globular clusters. The cool BHB stars required are present 
in the G1 color-magnitude diagram. 
The metallicity deduced for G1 is like that of the Galactic
globular 47 Tucanae, whose color-magnitude diagram is also very
similar. The age inferred is comparable to that of the Galactic halo.

Provided the elevated flux level below 2500\angs\ in the G1 spectrum 
is not due to scattered light,
the extremely hot EHB stars are also required, in numbers comparable to those 
of the cool BHB stars.
The EHB stars are too faint for detection in
the red CMD of G1, but in the Galaxy, 
they outnumber cool BHB's in both the metal-rich
open cluster \ngc 6791 (Liebert \etal) %\citep{lie94}
and in the globular cluster \ocen\ \citep{dcr00}.

Our composite spectrum analysis is limited in considering only 
a few stellar 
types, and in adopting but a single metallicity.
Nonetheless, we may draw several 
general conclusions.

\begin{itemize}

\item The near UV provides important additional constraints to
interpreting the spectra of remote composite systems. However, 
the age-metallicity degeneracy remains at some level even in the near UV.

\item The mid-UV spectral region can be modeled reliably by 
theoretical spectral calculations, but up-to-date models and 
line lists must be used. As demonstrated by PDR01, 
models of stellar photospheres should have convective overshoot 
turned off. Line lists must incorporate revised {\it gf}-values
and atomic lines not yet identified in the laboratory, and so 
must be tested by 
comparison with observed spectra of stars spanning the 
relevant range of spectral type and metallicity.

\item Blue horizontal branch stars must be included in composite spectra,
now that they are observed in many metal-rich systems in the Galaxy as 
well as in M31 G1 itself.

\item When BHB stars are 
included, the age of G1 does not appear to differ
substantially from that of Galactic globular clusters.
Ages of Andromeda globulars will be underestimated
from the Balmer lines whenever BHB stars are present but not accounted for.

\end{itemize}

These simple coadditions should be extended in several ways. 

\begin{itemize}

\item 
Main sequence, turnoff, subgiant, and red giant weighting factors should be
calculated from stellar isochrones including light-element
enhancements, and compared directly to in-depth star counts of the 47 Tuc
CMD. Because several Galactic globulars have rather different CMD's
within and outside their cores, attention should be paid to the spatial region
probed by the CMD. 

\item Plausible distributions of BHB and blue stragglers should
be included, both separately and in the same model, since \ocen\ and many 
other Galactic globulars show both \citep[e.g.,][]
{fer02}.

\item A variable overall metallicity should also be considered.
The relative abundances of critical 
elements such as magnesium, calcium, and nitrogen should be varied as well.

\item A large grid of theoretical
model spectra must be calculated to allow accurate interpolation for
all of these possible components.

\end{itemize}

We plan to continue this work under 
our three-year Hubble Treasury program GO-9455.
The mid-UV line list of PDR01 will first be improved, using updated 
laboratory measurements and new comparisons against 
high-quality mid-UV spectra to be taken in Cycles 11 and 12 
of MSTO stars, giants, and BHB/EHB stars of the field, 
and blue-straggler and turnoff stars in open clusters.
This should provide good mid-UV fits 
for stars of solar metallicity and higher, and for stars hotter and 
cooler than the Sun. It will also provide a more extensive
set of empirical mid-UV stellar spectral templates.
We will then calculate complete grids of theoretical stellar mid-UV spectra,
and generate composite spectra using dozens of stellar models rather
than the six considered here. Their relative weights will be derived from
isochrones calculated both with and without enhanced abundances of
magnesium, nitrogen, and other light elements. They will be tested 
against STIS spectra to be taken of G1 and four
other M31 globular clusters. 

All of our STIS observations are being made public as
they are taken, and the grids of theoretical stellar and composite 
spectra will be made public as they are completed. At that point, it
should be possible for anyone to judge from mid-UV and optical
spectra to what extent BHB stars might be contributing to the strong
\Hbeta\ indexes seen in most metal-rich M31 globular clusters, and to what
extent their ages are affected.

\acknowledgements 
We thank J. Fulbright of the Dominion Astrophysical Observatory for providing
the optical spectrum of \hd 106516, and M. Bessell for communicating 
and analyzing theoretical OH and CH transition probabilities. R. Schiavon and 
J. Rose offered many suggestions to improve the manuscript.
R.C.P. acknowledges support through grants to Astrophysical Advances 
for work on ultraviolet spectra from the Space Telescope Science Institute 
(GO-7395, GO-7402, AR-8371, and GO-9455)
and from NASA (ADP contract S-92512-Z and LTSA contract NAS5-02052), and 
for work on blue horizontal branch stars from the National Science 
Foundation (AST-9900582 and AST-0098725). BWC acknowledges support from 
NSF grant AST-9988156. RWO is partially supported by LTSA grant NAG5-6403.
RTR is partially supported by STScI grant GO-8709.

\figcaption{Plots are shown comparing observed
(heavy line) and
calculated (light line) spectra for the M31 globular cluster
G1 and a comparison star, \hd 106516. The left panel shows the mid-UV region,
and the right panel the optical. Wavelengths in air are given at the bottom.
The tick marks are separated
by 10\% of the full scale of each plot.
The uppermost panel shows a match about as good as that for the comparison
star. As described in the text, this match is achieved by adding blue horizontal
branch stars to the composite spectrum.}

\figcaption{As for Fig.\ 1.}

\figcaption{Plots are shown of several of the spectra calculated
from individual stellar models that were used to generate the composite
theoretical spectra of Fig.\ 1.}

\clearpage

\begin{figure}
%\epsscale{.79} %%commented out by submitter, file is 1
\epsscale{.85}  %%85% new size
\plotone{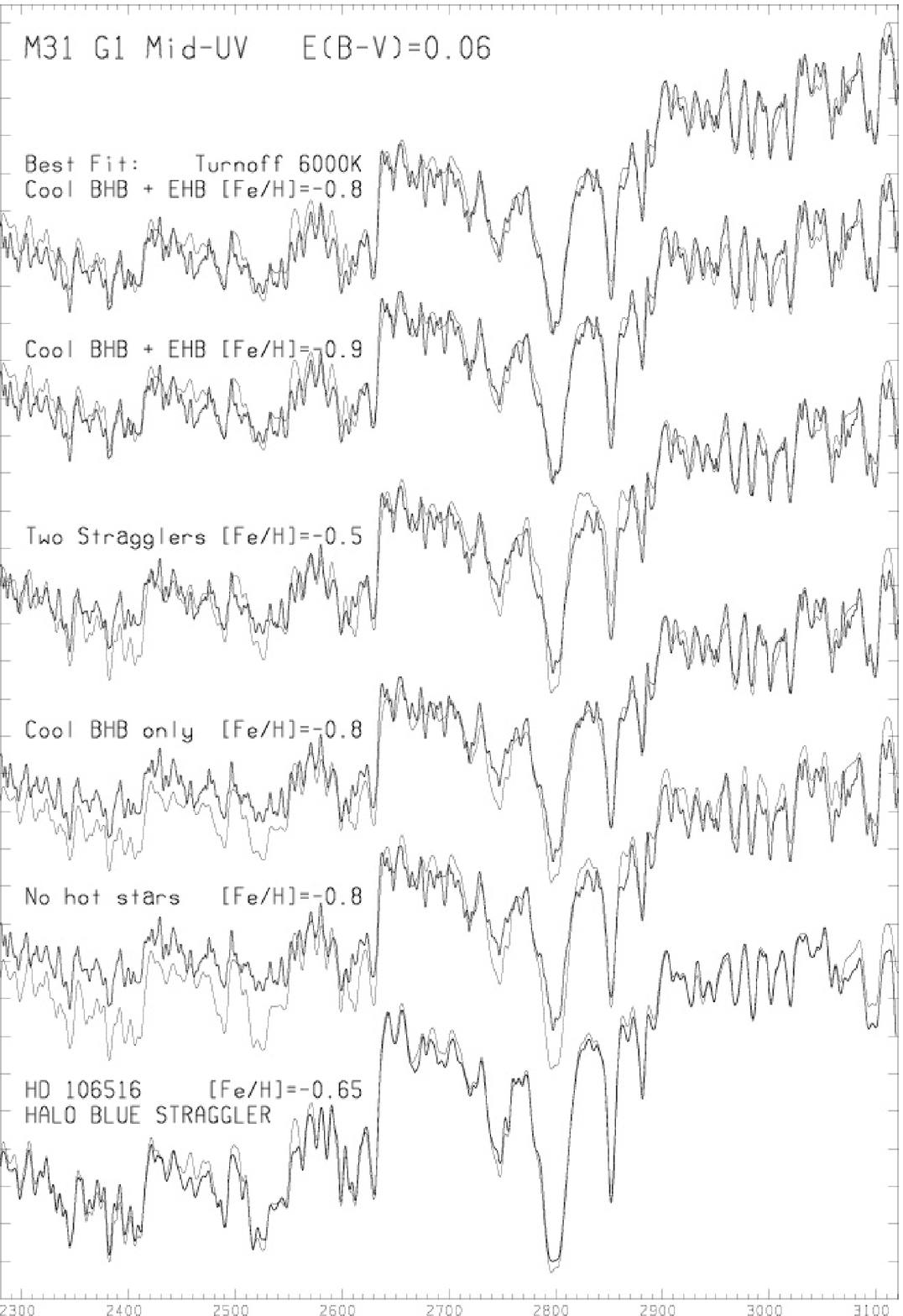}
\end{figure}
\begin{figure}
%\epsscale{1.118} %%defined by submitter, commented out by arXiv
\epsscale{.95}   %% 85% new size
\plotone{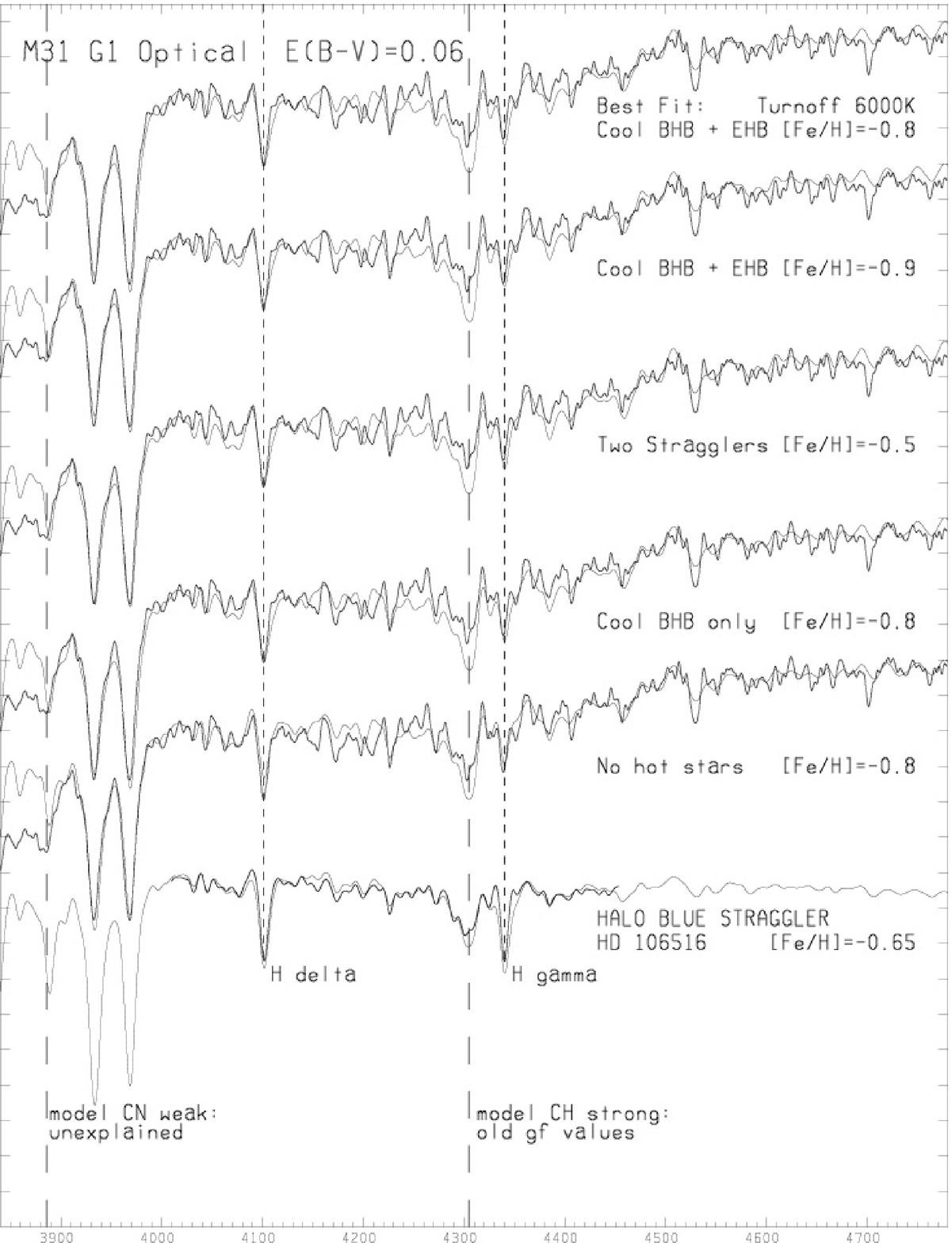}
\end{figure}

\begin{figure}
%\epsscale{.79} %%commented out by submitter, file is 1
\epsscale{.85}  %%85% new size
\plotone{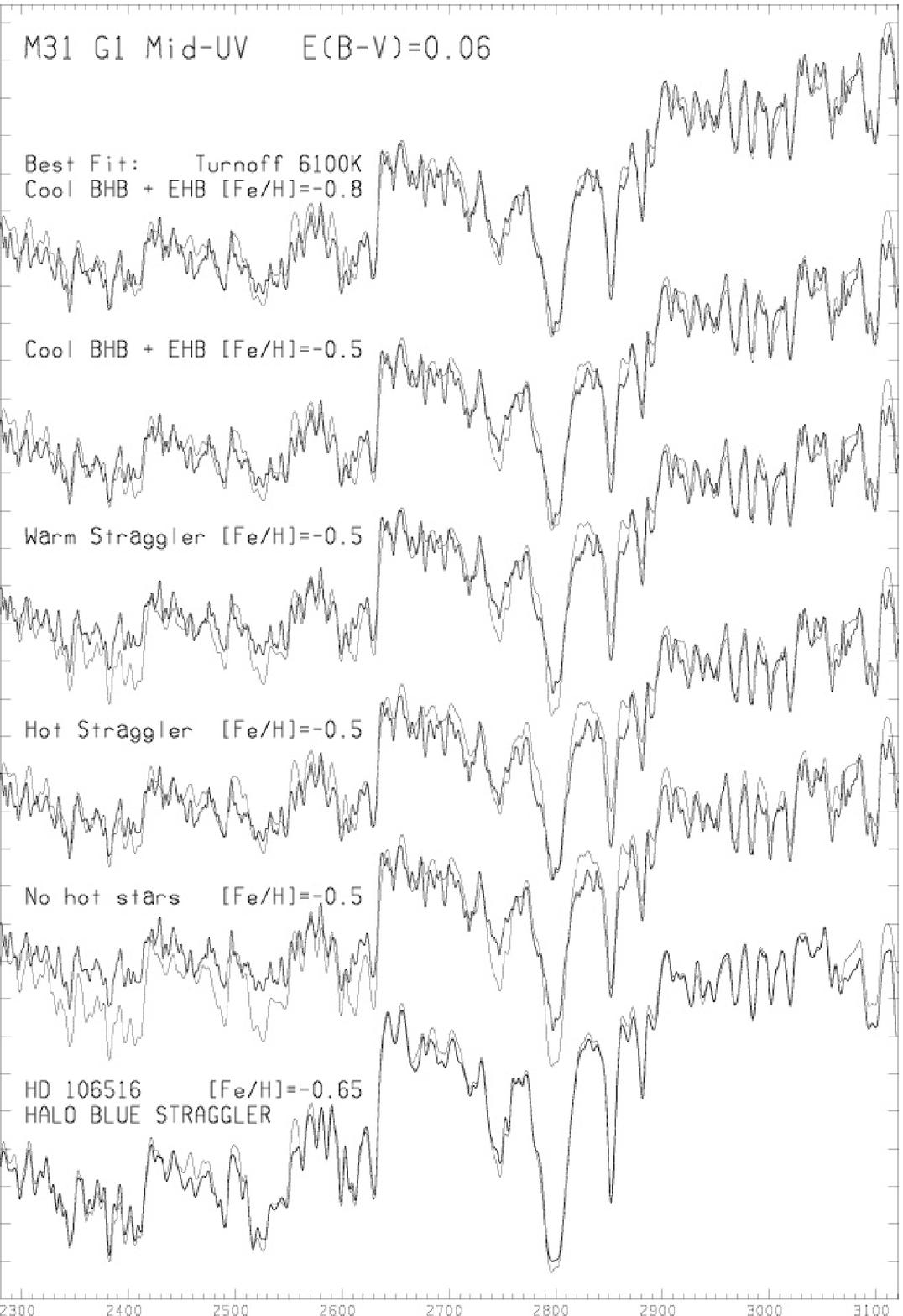}
\end{figure}
\begin{figure}
%\epsscale{1.118} %%defined by submitter, commented out by arXiv
\epsscale{.95}   %% 85% new size
\plotone{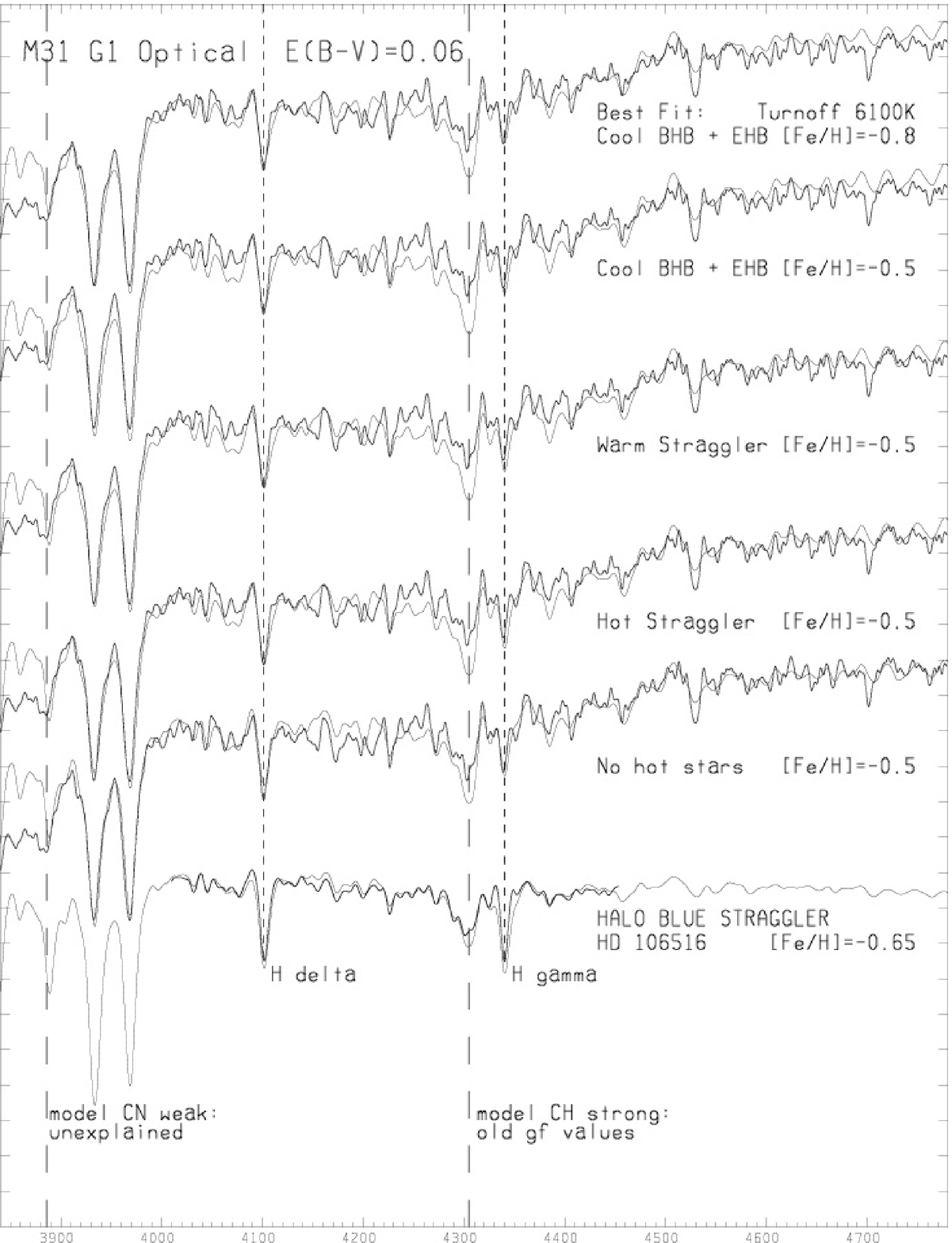}
\end{figure}

\begin{figure}
%\epsscale{.79} %%commented out by submitter, file is 1
\epsscale{.85}  %%85% new size
\plotone{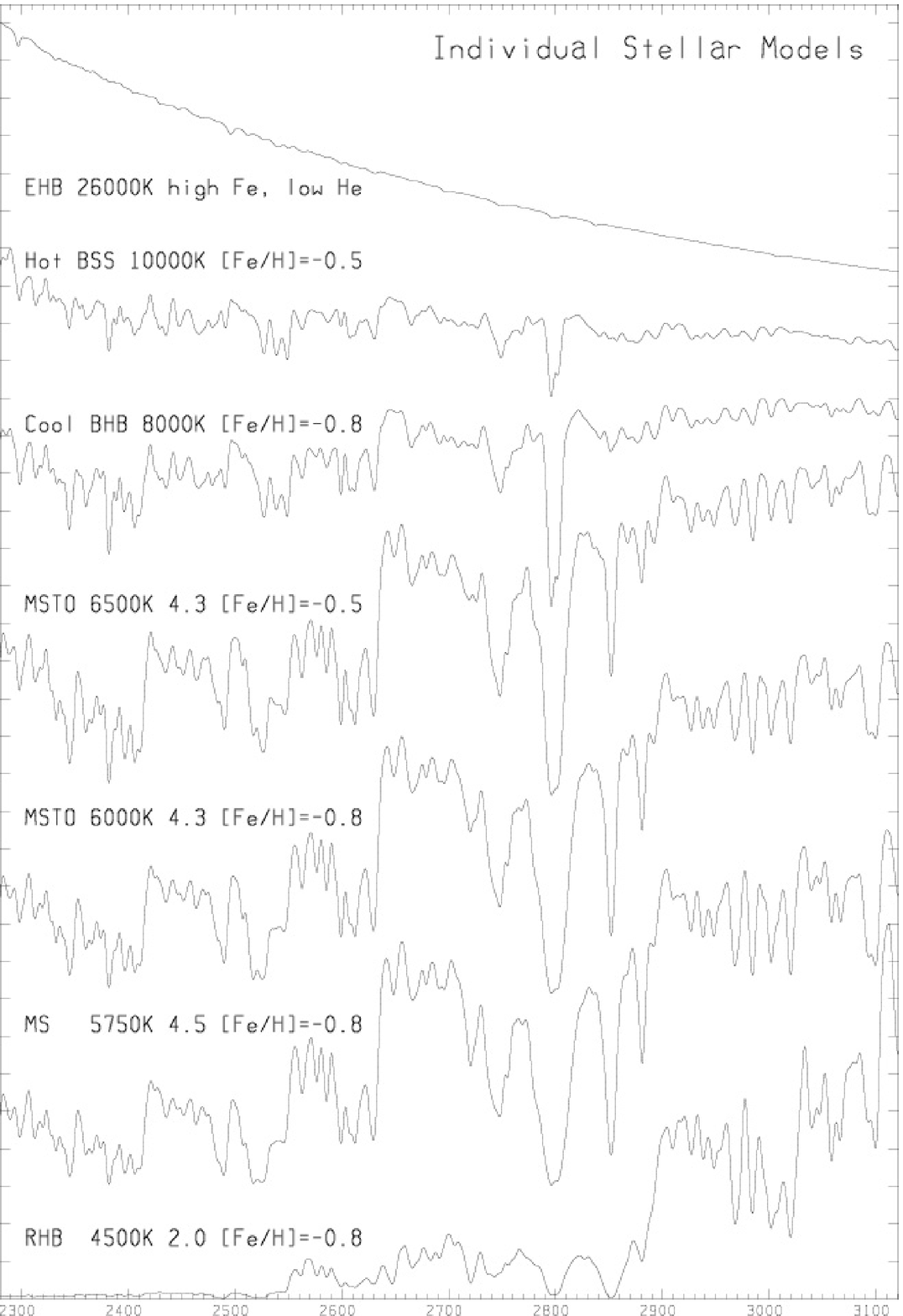}
\end{figure}
\begin{figure}
%\epsscale{1.118} %%defined by submitter, commented out by arXiv
\epsscale{.95}   %% 85% new size
\plotone{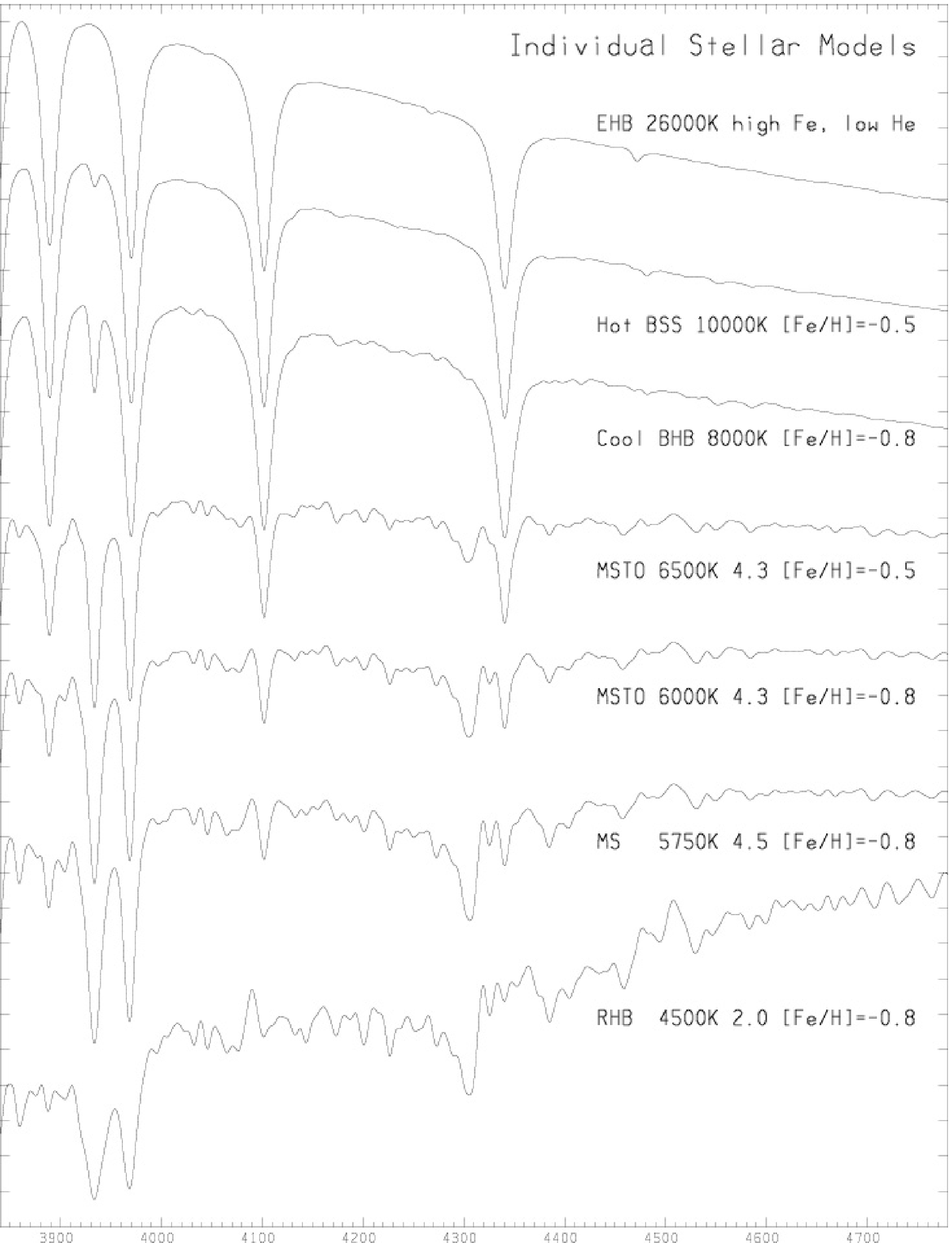}
\end{figure}

\end{document}